# Atomic-scale analyses of Nb₃Sn on Nb prepared by vapor-diffusion for superconducting radiofrequency cavity applications: A correlative study

Jaeyel Lee[1,2], Sam Posen[2], Zugang Mao[1], Yulia Trenikhina[2], Kai He[1,3], Daniel L Hall[4], Matthias Liepe[4], David N Seidman[1,5]

[1]Department of Materials Science and Engineering, Northwestern University, Evanston, IL, 60208, USA

[2]Fermi National Accelerator Laboratory, Batavia, IL, 60510, USA

[3]Northwestern University Atomic and Nanoscale Characterization Experimental Center (NUANCE), Evanston, IL, 60208, USA

[4]Cornell Laboratory for Accelerator-Based Sciences and Education, Cornell University, Ithaca, NY, 14853, USA

[5]Northwestern University Center for Atomic Probe Tomography (NUCAPT), Evanston, IL, 60208, USA

**Abstract**

We report on atomic-scale analyses of the microstructure of an Nb₃Sn coating on Nb, prepared by a vapor diffusion process for superconducting radiofrequency (SRF) cavity applications using transmission electron microscopy (TEM), electron backscatter diffraction (EBSD) and first-principles calculations. Epitaxial growth of Nb₃Sn on a Nb substrate is found and four types of orientation relationships (ORs) at the Nb₃Sn/Nb interface are identified by electron diffraction or high-resolution scanning transmission electron microscopy (HR-STEM) analyses. Thin Nb₃Sn grains are observed in regions with a low Sn flux and they have a specific OR: Nb₃Sn $(1\bar{2}0)$//Nb $(\bar{1}11)$ and Nb₃Sn $(002)$//Nb $(0\bar{1}1)$. The Nb₃Sn/Nb interface of thin grains has a large lattice mismatch, 12.3%, between Nb $(0\bar{1}1)$ and Nb₃Sn $(002)$ and a high density of misfit dislocations as observed by HR-STEM. Based on our microstructural analyses of the thin grains, we conclude that the thin regions are probably a result of a slow interfacial migration with this particular OR. The Sn-deficient regions are seen to form initially at the Nb$_3$Sn/Nb interface and remain in the grains due to the slow diffusion of Sn in bulk Nb$_3$Sn. The formation of Sn-deficient regions and the effects of interfacial energies on the formation of Sn-deficient regions at different interfaces are estimated by first-principles calculations. The finding of ORs at the Nb$_3$Sn/Nb interface provides important information about the formation of imperfections in Nb$_3$Sn coatings, such as large thin-regions and Sn-deficient regions, which are critical to the performance of Nb$_3$Sn SRF cavities for accelerators.

1. Introduction

$Nb_3Sn$ is an A15-type superconductor, which has been actively studied and used in superconducting wire applications [1]. A number of studies have employed $Nb_3Sn$ coatings on Nb for superconducting radiofrequency (SRF) cavity applications and these studies were motivated by the high critical temperature ($T_c$) and quality factor ($Q_0$) of this superconductor at a given temperature, compared to Nb [2-4]; $Q_0$ is defined by the surface resistance ($R_s$) and the geometric factor (G) of a cavity as $G/R_s$. $Nb_3Sn$ has lower surface resistance and a higher $T_c$ than the Nb [5, 6]. Hence, a $Nb_3Sn$ SRF cavity is a promising candidate to replace the current Nb SRF cavities for accelerator applications. Recently, studies at Cornell [7, 8] reported a high Q-factor of approximately $10^{10}$ at 4.2 K, with a maximum accelerating electric field gradient up to 17 MV/m for ~2 μm thick $Nb_3Sn$ coatings on Nb, prepared by a vapor diffusion process. Active research in vapor diffusion $Nb_3Sn$ films is now on-going at Fermilab, Cornell University and Jefferson Laboratory [9].

$Nb_3Sn$-coated cavities have been seen to quench superconductivity in the 14-17 MV/m range and some cavities still display a Q-slope, the increase of surface resistance as a function of accelerating field, Fig.1. The surface magnetic field at which the quench occurs, ~70 mT, is, however, significantly lower than the superheating field of $Nb_3Sn$ at ~400 mT, the ultimate limit predicted by the theory for an RF superconductor with an ideal surface [10-12]. These limits have been suggested to be a consequence of imperfections in the $Nb_3Sn$ coatings [13], including surface roughness, thin regions, Sn-deficient regions, grain boundaries, and surface chemistry.

Recent microstructural analyses have focused on two imperfections of $Nb_3Sn$ coatings on Nb, which are anticipated to have significant detrimental effects on the performance of $Nb_3Sn$ coated cavities: patchy regions with extremely thin grains [14-16] and Sn-deficient regions [14, 17]. In a joint study between Fermilab and Cornell, coupons were cut out from a cavity, which exhibited a substantially degraded $Q_0$, see Q vs E curve for cavity ERL1-5 displayed in Figure 1, which was linked to localized areas with high surface resistance values. Electron microscopy was performed on coupons cut from the wall of the cavity, which revealed that areas with a high surface-resistance contained grains with an abnormally large transverse length and a thickness of only ~100 nm (compared to a normal value of ~2 μm) [9, 14, 18]. The penetration depth of magnetic fields in $Nb_3Sn$ is ~111 nm [12] and $Nb_3Sn$ coatings require a thickness of at least ~500 nm to shield the magnetic field effectively and avoid excessive dissipation from poorly superconducting Nb/Sn compounds in the $Nb_3Sn$/Nb interface [14]. The formation of thin grains (< ~500 nm) in the $Nb_3Sn$ coating has been reported to be affected by a number of factors including the supply of Sn [19], pre-anodization of the Nb substrate [18], and crystallographic orientation of Nb [15]. There is some evidence that the texture and nucleation of $Nb_3Sn$ grains could play a role but the detailed mechanism and origin of the formation of thin grains are still not understood [15]. Also, Sn-deficient regions are one of the primary concerns for $Nb_3Sn$ superconducting radiofrequency cavities [17]. The proportion of Sn in $Nb_{3+x}Sn_{1-x}$ ranges from ~17 to 26 at.%, Fig. 1 [1, 20], and the $T_c$ of $Nb_3Sn$ varies from 6 K at ~17 at.% Sn to 18.3 K at ~26 at.%. Therefore, Sn-deficient regions with 17-19 at.% Sn can decrease the $T_c$ of $Nb_3Sn$ coatings to below that of Nb, 9 K. The formation of Sn-deficient regions is particularly undesirable near the top surface of $Nb_3Sn$ cavities where radiofrequency currents flow. The growth mechanism and compositional variation of $Nb_3Sn$ have been rigorously investigated in $Nb_3Sn$ samples prepared for superconducting wire applications by a solid-diffusion process using Cu-Sn and Nb diffusion couples, the bronze process [1, 21-25]. The compositional variation in $Nb_3Sn$ prepared by solid-diffusion was investigated utilizing a composition gradient between a Sn-rich phase ($Nb_6Sn_5$) and bulk Nb, originating from the diffusion process [21, 25-27]. Low levels of Sn in $Nb_3Sn$ have been attributed to the small formation energy of Nb antisite point-defects, Nb sitting on Sn sites [26]. There is, however, little information about how Sn-deficient regions are formed during vapor diffusion preparation of a $Nb_3Sn$ coating on Nb or about the critical role(s) played by the $Nb_3Sn$/Nb interface in $Nb_3Sn$ grain-growth and the formation of thin grains. The objective of the current studies is to investigate the origin of the formation of thin grains and Sn-deficient regions in $Nb_3Sn$ coatings on Nb prepared by the vapor-diffusion process and the role of $Nb_3Sn$/Nb interfaces on the formation of imperfections in general.

Herein, we report on atomic-scale analyses of $Nb_3Sn$ coatings on Nb using transmission electron microscopy (TEM), electron backscatter diffraction (EBSD) and first-principle calculations. In particular, we find that orientation relationships (ORs) at the $Nb_3Sn$/Nb interface are correlated with the formation of thin-grains and Sn-deficient regions. Through the use of electron diffraction and high-resolution scanning transmission electron microscopy (STEM), we identify four types of orientation relationships (ORs) for $Nb_3Sn$/Nb heterophase interfaces. Notably, large thin $Nb_3Sn$ grains are highly correlated with a certain grain orientation-relationship, specifically $Nb_3Sn$ $(1\bar{2}0)$//Nb $(\bar{1}11)$ and $Nb_3Sn$ $(002)$//Nb $(0\bar{1}1)$. Also, first-principles calculations of $Nb_3Sn$/Nb interfaces demonstrate that the formation of Sn-deficient $Nb_3Sn$ close to Nb substrates is attributed to the smaller interfacial free energy of Sn-deficient $Nb_3Sn$/Nb interfaces compared to stoichiometric $Nb_3Sn$/Nb. Our findings provide evidence for understanding the formation of imperfections in $Nb_3Sn$ coatings on Nb, which may be used to improve the performance of $Nb_3Sn$ SRF cavities.

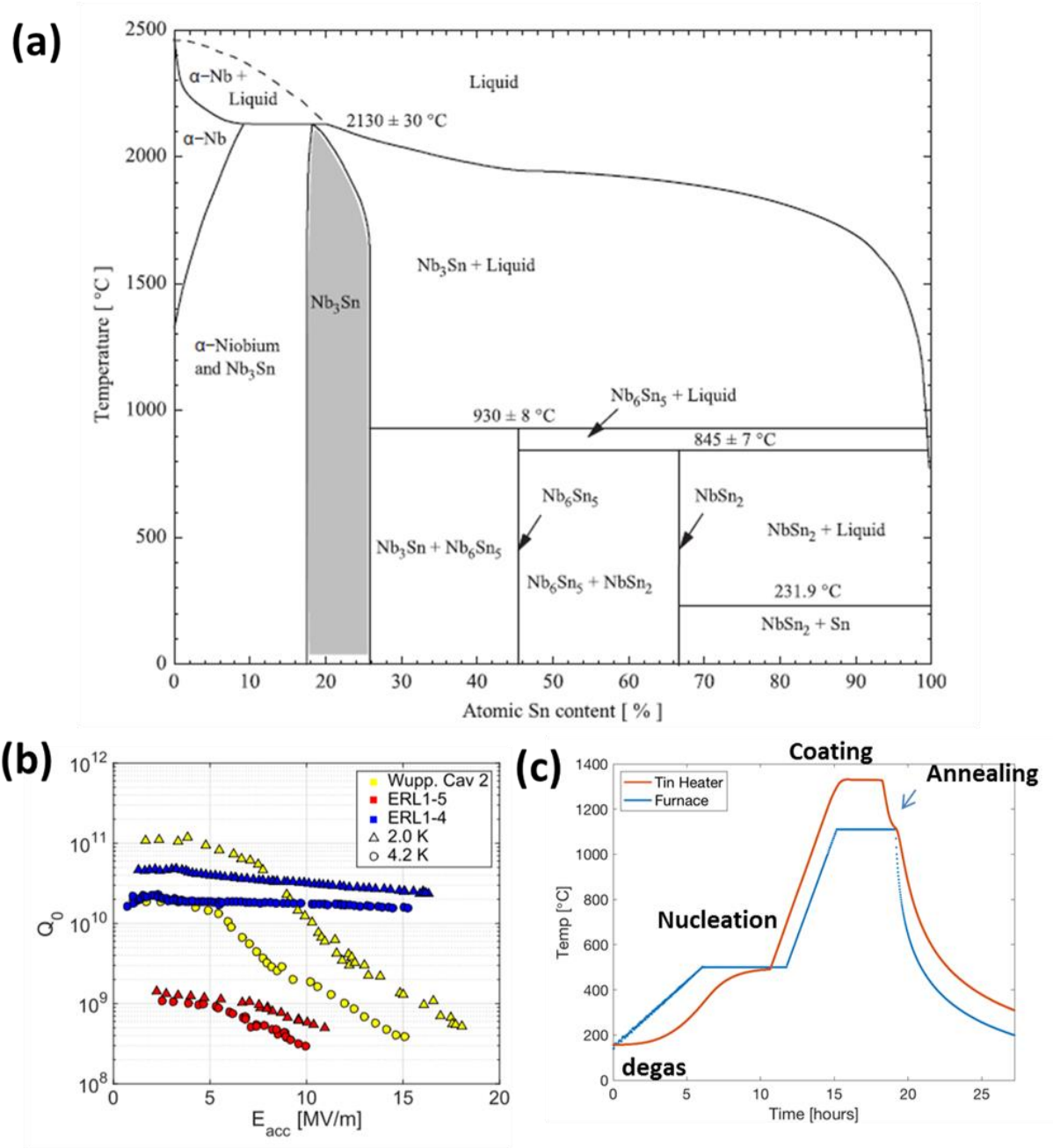

Figure 1. (a) Nb-Sn binary phase diagram [1, 20]. (b) Quality factor ($Q_o$) vs electric field (E) curves for different ~2 μm thick $Nb_3Sn$ coated superconducting radiofrequency (SRF) cavities [14]. $Nb_3Sn$ coated SRF cavity of ERL 1-4 displays a high $Q_o$-factor up to 15-17 MV/m, which is close to the state-of-art performance for a $Nb_3Sn$ SRF cavity. ERL 1-5 exhibits a low $Q_o$-factor, that is, a high surface resistance. Thin-grain regions were widely observed in the $Nb_3Sn$ SRF cavity of ERL 1-5, with a strong correlation

observed between the surface resistance of Nb$_3$Sn and microstructures. (c) Temperature profile of substrate and tin source during the coating procedure [9, 18].

2. Methodologies

2.1 Experimental procedures

Nb$_3$Sn coatings were prepared at the Fermi National Accelerator Laboratory (FNAL) and Cornell University by a vapor diffusion process, Fig. 1(c); for development of the process see Refs. [2, 3, 9, 10, 28, 29]. Niobium samples, tin source, and a SnCl$_2$ nucleating agent are placed in a vacuum furnace and heated to 500 ˚C to create nucleation sites of Nb$_3$Sn on the Nb surface. The furnace temperature was then raised to 1100 ˚C for 3.5 h to permit a Nb$_3$Sn coating to form on the Nb surface. The tin source was maintained at 1200 ºC, so that sufficient Sn could continue to be provided to the top surface of Nb$_3$Sn, which subsequently diffuses into the Nb sample.

In our studies of nucleation of Nb$_3$Sn, only a SnCl$_2$ nucleating agent was placed in the furnace, without Sn, and then heated to 500 ˚C to induce nucleation. Next, the temperature was increased to 1100 ˚C and then cooled immediately after reaching 1100 ˚C.

The samples were systematically characterized by scanning electron microscopy (SEM), transmission electron microscopy (TEM) and electron backscatter diffraction (EBSD). A 600i Nanolab Helios Focused Ion-Beam (FIB) microscope was employed to prepare cross-sectional samples for TEM. The samples were thinned employing a 30 kV Ga+ ion-beam at 27 pA, and fine-polished using 5 kV Ga+ ions at 47 pA. A Hitachi HD-8100 was used for the bright-field (BF) TEM imaging and electron diffraction analyses, and a Hitachi HD-2300 was used for STEM and energy dispersive spectroscopy (EDS). High-resolution TEM images were recorded using a JEOL Grand ARM-300 and high-resolution STEM images were acquired using a JEOL aberration-corrected Grand ARM-200. The Gatan Micrograph Suite version 2.11 was used to analyze and process the images. EBSD data was collected using a FEI Quanta field-emission gun (FEG) SEM equipped with an HKL Nordlys S camera utilizing Oxford AZtec EBSD software. The surfaces of samples were placed at a 10 mm working distance and a 30 kV electron beam with an incident angle of 20$^o$ was used for accumulating EBSD patterns. For transmission (t-) EBSD, an EM-Tec TE3 t-EBSD holder was utilized to place the surface of a TEM sample at an incident beam angle of 20$^o$.

### 2.2 Computational details

The first-principles calculations in this research employed the plane-wave pseudopotential total energy method as implemented in the Vienna ab initio simulation package (VASP) [30]. We used projector augmented wave (PAW) potentials [31] and the generalized gradient approximation (GGA), developed by Perdew-Burke-Ernzerhof (PBE) [32] for exchange-correlation. Unless otherwise specified, all structures are fully relaxed with respect to volume, as well as all cell-internal atomic coordinates. We considered carefully the convergence of results with respect to energy cutoff and k-points. A plane-wave basis set was used with an energy cutoff of 600 eV to represent the Kohn-Sham wave functions. A summation over the Brillouin zone for the bulk structures was performed on a 12×12×12 Monkhorst-pack k-point mesh for all calculations and a magnetic spin-polarized method was applied for all calculations. The calculated lattice parameters of Nb and $Nb_3Sn$ are 3.324 and 5.332 Å, respectively, which are in excellent agreement with the experimental results, 3.300 Å for Nb [33] and 5.289 Å for $Nb_3Sn$ [34]. Both 2×2×2 and 3×3×3 supercells were used to determine the vacancy formation energies, antisite energies, and lattice substitutional energy calculations.

### 3. Experimental results

#### *3.1 Transmission electron microscopy analysis of nucleated $Nb_3Sn$ grains*

As a first step, we characterized nucleated $Nb_3Sn$ grains in the early stages of the $Nb_3Sn$ coating process to observe the details of the initial grain growth. The SEM micrograph in Fig. 2 shows nucleated $Nb_3Sn$ grains on a Nb surface, with some displaying lateral growth with a flat morphology. Figure 2 displays three connected nucleated $Nb_3Sn$ grains with diameters of ~500 nm (n1), ~200 nm (n2), and ~100 nm (n3), respectively. Two of these nucleated Nb$_3$Sn grains were selected and cross-sectional TEM samples were prepared from them using a dual-beam FIB microscope. High-angle annular dark-field (HAADF) scanning transmission electron microscopy (STEM) images in Fig. 2 reveal that the thickness of the nucleated grain n1, was ~100 nm and that of n2 was ~60 nm. Energy dispersive spectroscopy (EDS) mapping of Nb Lα (2.17 keV) and Sn Lα (3.44 keV) lines performed in the STEM mode indicates that the concentration of Sn in the nucleated Nb$_3$Sn grain is 16-19 at.%, which is Sn-deficient compared to the nominal concentration of 25 at.% Sn in Nb$_3$Sn.

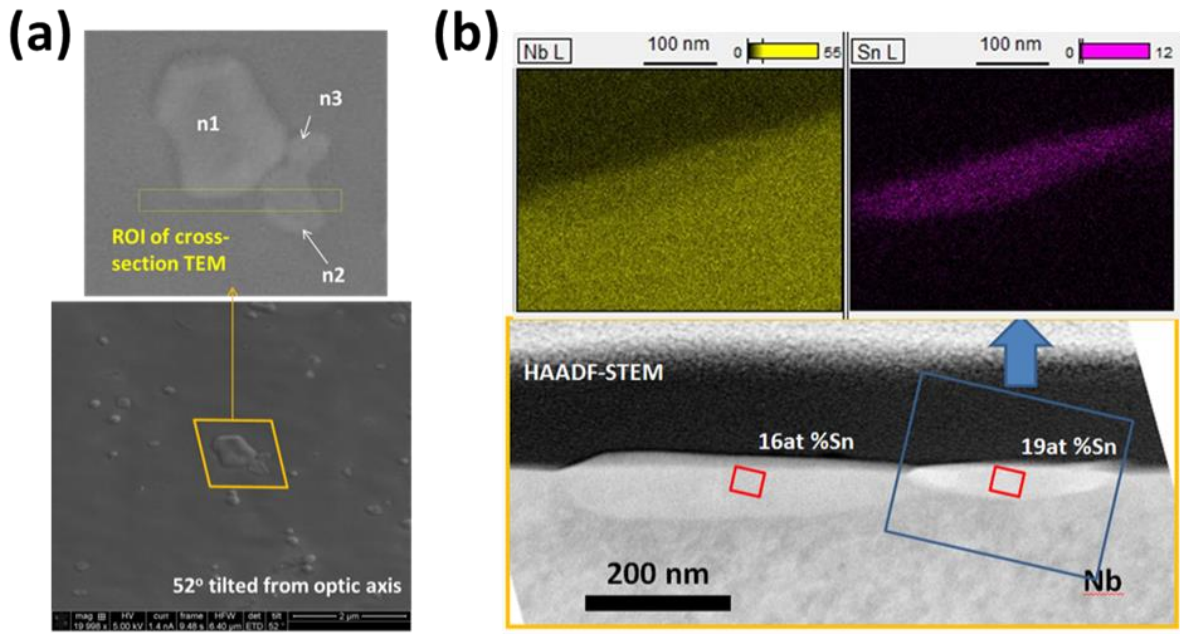

Figure 2. (a) Scanning electron microscope (SEM) images of nucleated Nb$_3$Sn grains prepared at Fermilab. Three nucleated grains are labeled n1, n2, and n3. The region for cross-sectional transmission electron microscopy (TEM) is indicated by a parallelogram. (b) high-angle annular dark-field scanning transmission electron microscope (HAADF-STEM) image and energy-dispersive spectroscopy (EDS) maps of Nb Lα (2.17 keV) and Sn Lα (3.44 keV) of the nucleated Nb$_3$Sn grains.

### 3.2 Transmission electron microscopy analyses of Nb$_3$Sn grains

The microstructures of the Nb$_3$Sn coating and the Nb$_3$Sn/Nb interfaces were analyzed employing bright-field transmission electron microscopy (BF-TEM), high-angle annular dark-field scanning transmission electron-microscopy (HAADF-STEM), EDS, and high-resolution-scanning/transmission electron-microscopy (HR-S/TEM) after the coating process was completed. The results displayed significant variations in the microstructures with respect to the net Sn-flux, which was estimated by taking the average thickness of the Nb$_3$Sn coating and the coating process time. It is possible to distinguish three different regions (Table 1): (1) abnormally large-grain regions for a high Sn-flux; (2) normal grain-size regions for a medium Sn-flux; and (3) patchy regions with thin grains for a low Sn-flux.

Table 1. Summary of the three regions of the Nb$_3$Sn coatings with different microstructures, which depend on the net Sn flux or growth rate.

|  | Net Sn flux [atoms/nm$^2$ min] | Average growth rate [nm/min] | Average thickness [μm] |
|---|---|---|---|
| Abnormally large-grain regions | 322 | 24 | ~5 |
| Normal-grain regions | 161 | 12 | ~2.5 |
| Patchy regions with thin-grains | 47 | 3.5 | ~0.7 |

#### 3.2.1  Abnormally large Nb$_3$Sn grain regions (average growth rate: 24 nm/min)

In the case of a high net Sn-flux (high average growth rate of 24 nm/min), we observed regions with abnormally large and irregularly shaped Nb$_3$Sn grains, SEM image of Fig 3(a), recorded using a secondary-electron (SE) detector. The regions have a rounded-shape and a size that varies from tens to hundreds of microns. Abnormally large grain size regions have bright contrast, compared to neighboring regions with normal-size grains, possibly due to topological effects from the rough $Nb_3Sn$ surface. HAADF-STEM imaging of a cross-section of $Nb_3Sn$/Nb, Fig 3(c), yields both the grain size and thickness, ~ 5 μm for each. STEM-EDS analyses of the abnormally large grains reveal a concentration of ~26 at.% Sn, which is similar to normal $Nb_3Sn$ grains, suggesting that the stoichiometry of the abnormally large grain regions is also $Nb_3Sn$. The round shape of the abnormally large grain regions and the irregular outline of large $Nb_3Sn$ grains in the region imply the abrupt formation of $Nb_3Sn$ phases from a liquid droplet of Sn, which is discussed further in section 4.1.

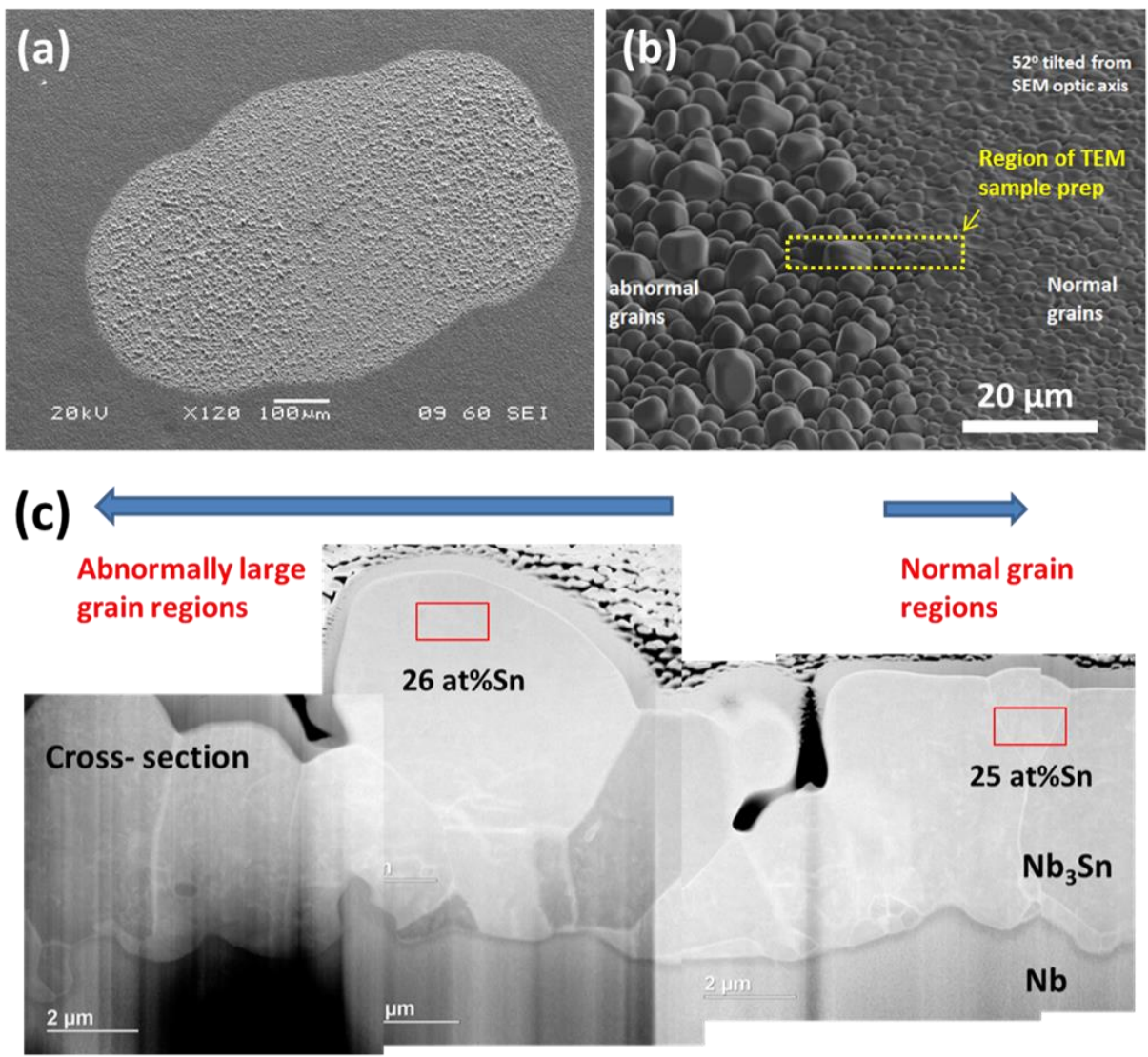

Figure 3. (a) Scanning electron microscope (SEM) image of abnormally large grain regions of the $Nb_3Sn$ coating prepared at Fermilab, recorded using a secondary-electron (SE) detector. (b) SEM image of the boundary between abnormally large-grain regions and normal grain-regions tilted by 52˚ from the SEM's optic axis. (c) HAADF-STEM image of a cross-section of $Nb_3Sn$/Nb across regions with normal grains and abnormally large grains. The composition of one of the abnormally large grains is estimated using STEM-EDS analyses exhibiting similar values (26 at.% Sn compared to 25 at.% Sn in the normal grains).

### 3.2.2  *Normal $Nb_3Sn$ grain regions (average growth rate: 12 nm/min)*

The HAADF-STEM image in Fig. 4(c) demonstrates that the $Nb_3Sn$ coatings on Nb are approximately 2.5 ± 1.0 μm thick, with its surface features revealed in the SE (secondary electron)-SEM image, Fig. 4(a). The sample in the SEM image was tilted by 52˚ from the SEM's optic-axis to display the granular roughness of the surface. Utilizing SEM imaging, the average grain size for two samples, A5 and A10, is 2.0 ± 0.6 μm. Quantitative analyses of the $Nb_3Sn$ coatings employing STEM-EDS reveals the presence of Sn-deficient regions. Most of these regions are close to the Nb substrate, although they are occasionally observed in the middle of $Nb_3Sn$ grains, Fig. 4(d). Quantification of the composition of two Sn-deficient regions (numbers 1 and 3) is ≈19 at.% Sn compared to the anticipated value of 25 at.% Sn for stoichiometric $Nb_3Sn$.

Next, we analyzed the $Nb_3Sn$/Nb interfaces utilizing BF-TEM, HR-TEM and electron diffraction. The results yield orientation relationships (ORs) between $Nb_3Sn$ and Nb at the $Nb_3Sn$/Nb interface. Fig 5(a) displays a HR-TEM image of the $Nb_3Sn$/Nb interface in normal $Nb_3Sn$ grain regions. The HR-TEM image was recorded for a Nb [$\bar{1}11$] zone-axis, which reveals the orientation relationships of the $Nb_3Sn$ grains with the Nb substrate. One grain (labeled Orientation A) has an orientation relationship with Nb of $Nb_3Sn$ ($1\bar{2}0$)//Nb ($\bar{1}11$) and $Nb_3Sn$ (002)//Nb ($1\bar{1}2$), while another grain (labeled Orientation C) is Nb$_3$Sn (1$\bar{2}$0)//Nb ($\bar{1}$11) and Nb$_3$Sn (002)//Nb (0$\bar{1}$1). Owing to the orientation relationship of two grains at a Nb$_3$Sn/Nb interface a [1$\bar{2}$0] tilt grain-boundary with 29° tilt-angle is formed. The HAADF-STEM image in Fig. 5(b) exhibits another OR at the Nb$_3$Sn/Nb interface for a Nb [011] zone axis. As seen in Fig. 5(c), the electron diffraction pattern of one of the grains at the Nb$_3$Sn/Nb interface (indicated by the red dotted circle in Fig. 5(b)) has the following ORs: Nb$_3$Sn (002)//Nb (011) and Nb$_3$Sn (130)//Nb (002) (termed Orientation D).

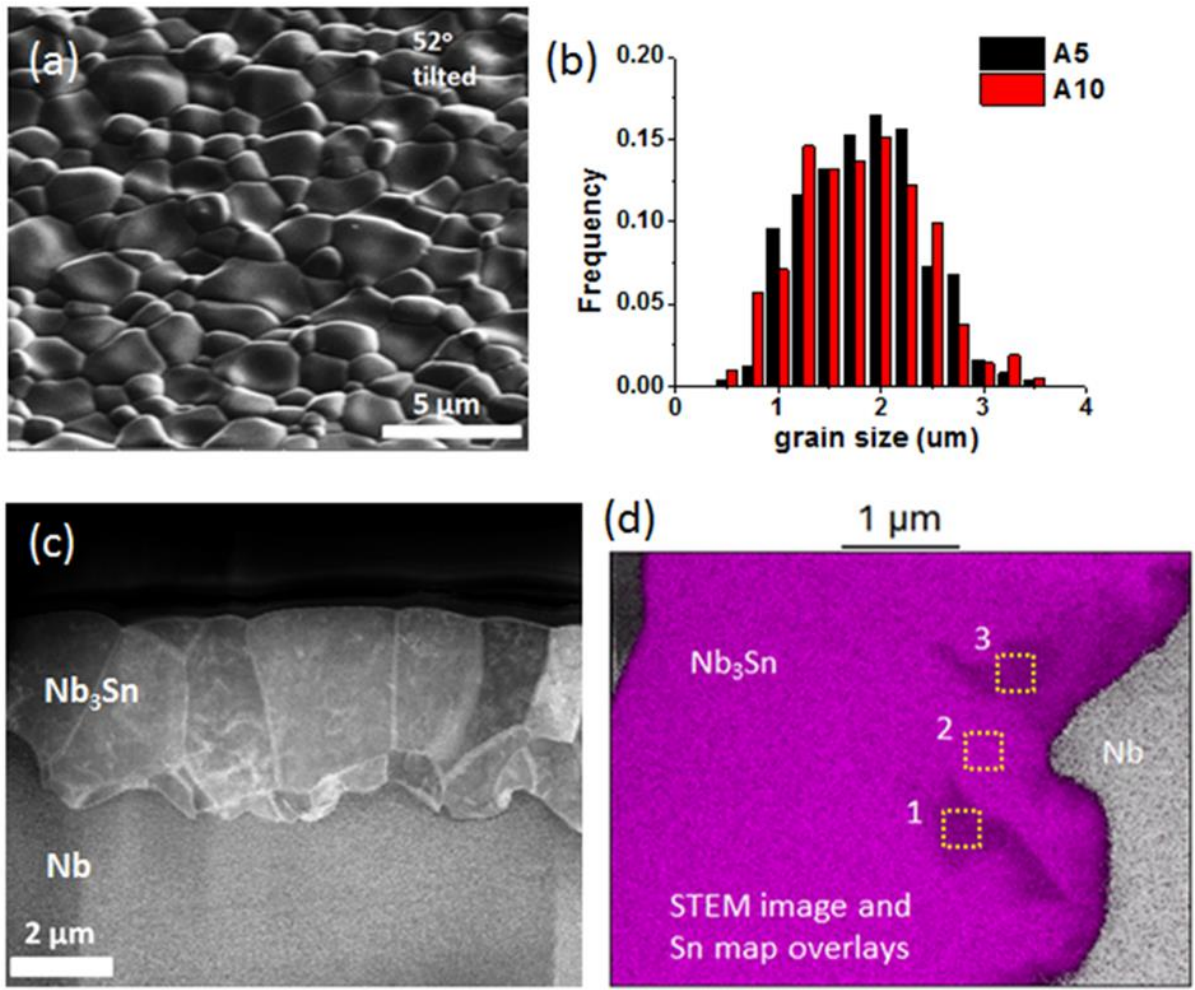

Figure 4. (a) Secondary-electron (SE) SEM image of a Nb$_3$Sn surface, which displays roughness due to the granularity of Nb$_3$Sn (the surface was tilted 52° from the SEM's optic axis). The samples were coated at Fermilab. (b) Histogram of grain sizes (diameters) of A5 and A10 samples with an average grain size

of 2.0 ± 0.6 μm. (c) A cross-sectional view of a Nb$_3$Sn film on Nb, obtained using HAADF-STEM. (d) An HAADF-STEM image and overlays of the STEM image and EDS Sn L map on it. Sn-deficient regions are observed near bulk Nb (denoted by 1 and 3). The tin concentration of region numbers 1 and 3 is 19 at.%, while that of region number 2 is 25 at.%.

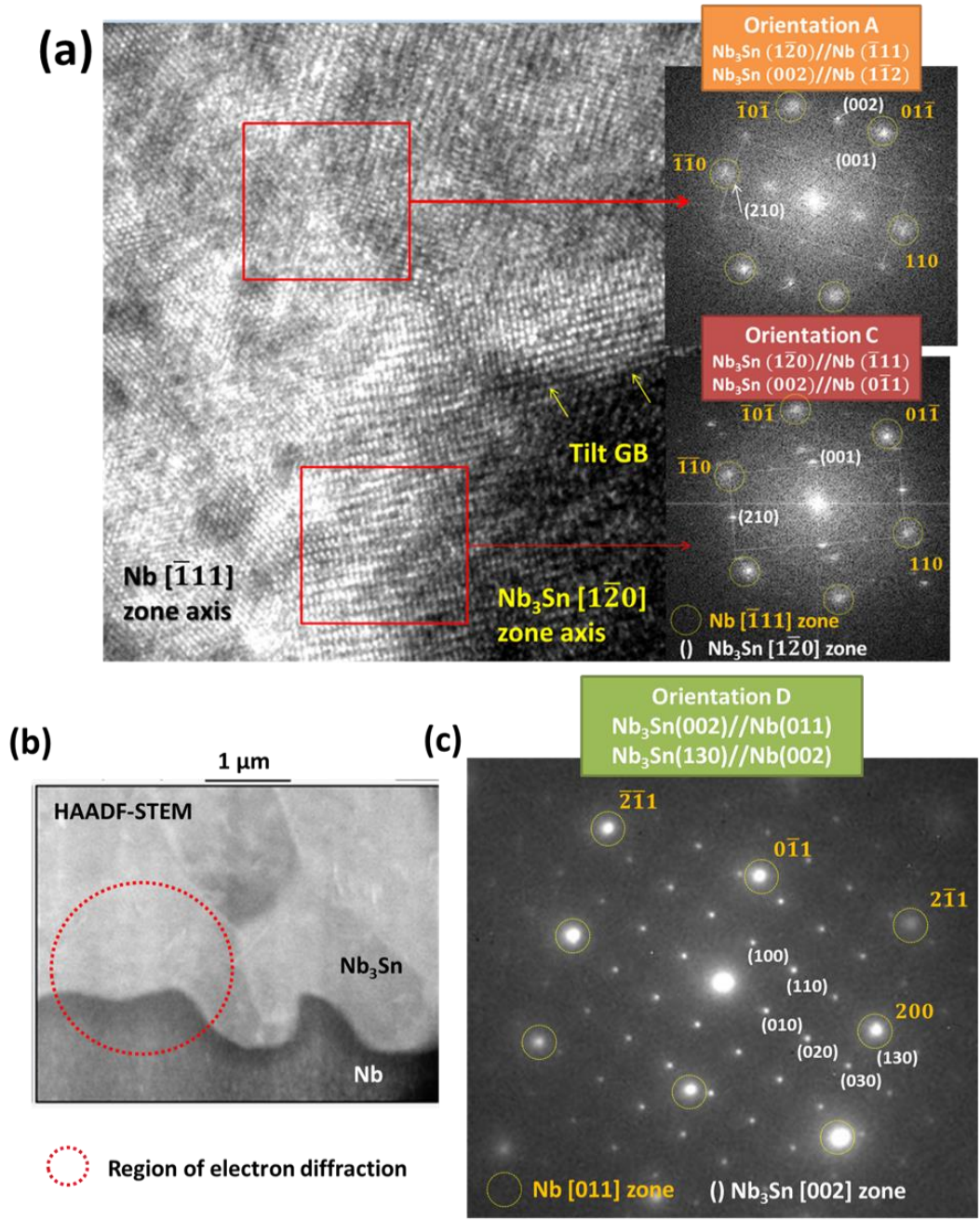

Figure 5. (a) HR-TEM image of Nb$_3$Sn/Nb for normal-grain regions, showing two grains grown epitaxially on Nb with their orientation relationships. The Nb$_3$Sn coating was prepared at Cornell University. The orientation relationships correspond to a 29$^o$/[1$\bar{2}$0] tilt grain-boundary. (b) HAADF-STEM image of a Nb$_3$Sn/Nb interface with a Nb [011] zone-axis. The region for the electron diffraction pattern is indicated by a red-dotted circle. The sample was prepared at Fermilab. (c) The electron diffraction pattern of the Nb$_3$Sn/Nb interface yields the following orientation relationships, Nb$_3$Sn (002)//Nb (011) and Nb$_3$Sn (130)//Nb (002), termed Orientation D.

*3.2.3   Patchy regions with thin Nb$_3$Sn grains (average growth rate: 5 nm/min)*

As previously reported [9, 14, 35], patchy regions including thin-grains with a large lateral grain size of up to many tens of microns form in regions with a low net Sn-flux of ~47 Sn atoms/nm$^2$ min (slow average growth rate of ~5 nm/min). The HAADF-STEM image in Fig. 6 presents one of the patchy regions with thin Nb$_3$Sn grains (grains numbers 1 and 5). The ORs of the Nb$_3$Sn/Nb interface, of the seven grains in the TEM sample, were analyzed by electron diffraction or HR-TEM, which are displayed in Fig. 6. The first grain to be analyzed is the relatively thick grain (grain number 3). Compositional variations within the regions of grains numbers 3 and 4 are examined by STEM-EDS, Fig. 7(c), which reveals that there are low levels of Sn in portions of the Nb$_3$Sn/Nb interface [17]. Atomically resolved HR-STEM images of the Nb$_3$Sn/Nb interface for grain number 3 were recorded using a JEOL aberration corrected Grand ARM-200, Fig. 7(a). This interface displays epitaxial growth of Nb$_3$Sn on Nb with Orientation A, Nb$_3$Sn (1$\bar{2}$0) //Nb ($\bar{1}$11) and Nb$_3$Sn (002) //Nb (1$\bar{1}$2) agreeing with the electron diffraction pattern of this interface in Fig. 7(b). The interplanar distance of Nb$_3$Sn (002) is 2.63 Å (JCPDS No. 04-017-6755), while the value for Nb (1$\bar{1}$2) is 2.70 Å (JCPDS No. 00-035-0789), which results in an ~2.8% lattice mismatch and produces a tensile strain in the Nb$_3$Sn grain along the [002] direction. There are additional misfit dislocations at the Nb$_3$Sn/Nb interface due to the lattice mismatch between Nb$_3$Sn and Nb. Another OR, Orientation B, is observed in grains numbers 2 and 4. As observed in the electron diffraction pattern of the Nb$_3$Sn/Nb interface of grain numbers 2 and 3, Fig. 8 (a), the (002) plane of Nb$_3$Sn grain number 2 is parallel to Nb $(23\bar{1})$, which implies that grain number 2 is tilted by 8.9° about the Nb$_3$Sn $[1\bar{2}0]$ axis from grain number 3 with Orientation A. The HR-TEM image of grain numbers 2 and 3, Fig. 8 (b), demonstrates that the Nb$_3$Sn grains form an 8.9°/$[1\bar{2}0]$ tilt grain boundary.

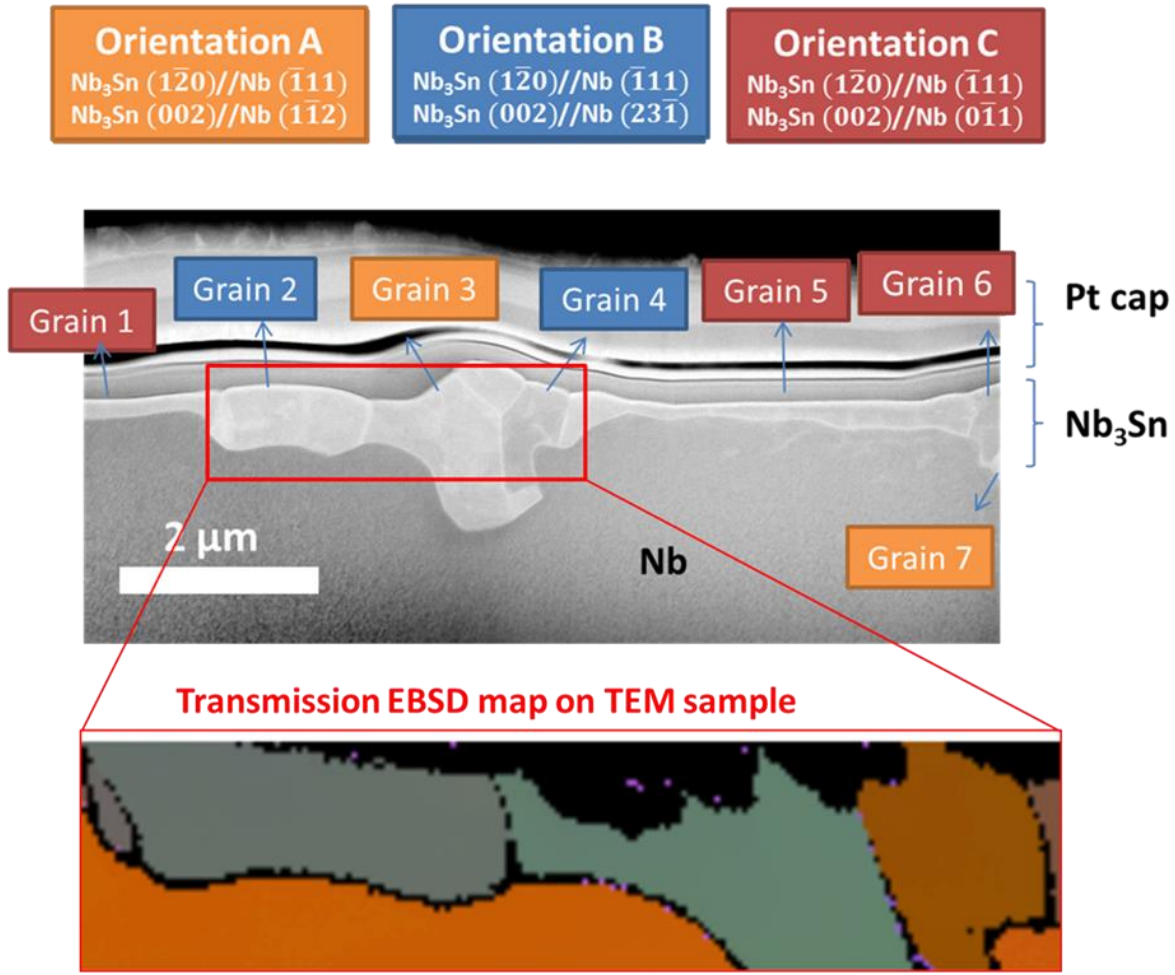

Figure 6. HAADF-STEM image of the patchy region with thin Nb$_3$Sn grains. Three types of orientation relationships (Orientations A, B, and C) are observed. Transmission electron back scattered diffraction (EBSD) mapping of four grains, grains numbers 2 to 5, was conducted to identify the misorientation axes and angles for Orientations A, B, and C. The Nb$_3$Sn film was coated at Fermilab.

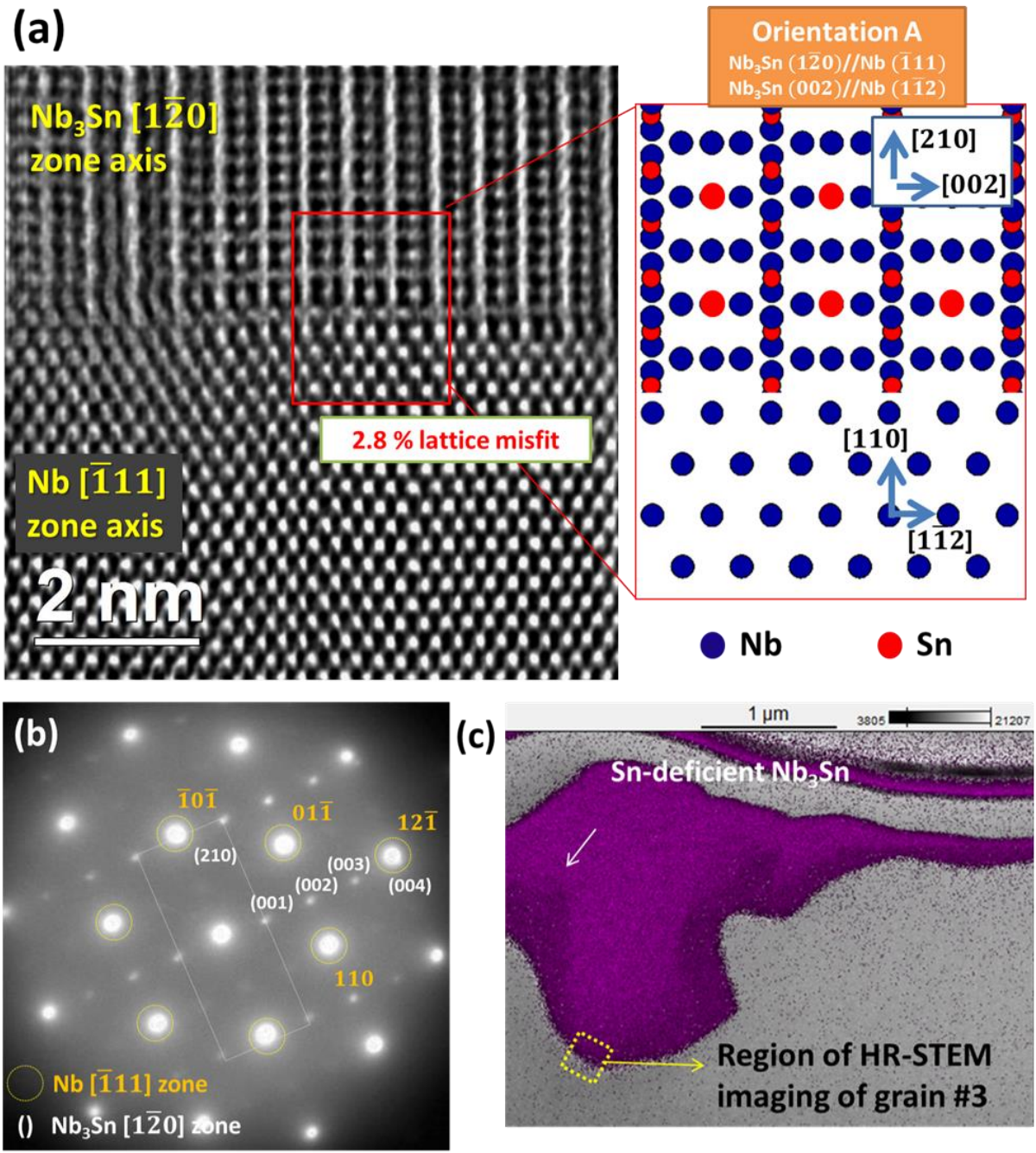

Figure 7. (a) Atomically-resolved HR-STEM image of the Nb$_3$Sn/Nb interfaces of grain number 3 and corresponding atomic configurations displaying epitaxial growth of Nb$_3$Sn on Nb with Orientation A, right-hand side. The solid-red circles are Sn atoms and the solid-blue circles are Nb atoms. (b) Corresponding electron diffraction pattern of the Nb$_3$Sn/Nb interface of grain number 3. (c) STEM-EDS

Sn Lα (3.44 keV) mapping of the region of grain numbers 3, 4 and 5. The area of the HR-STEM image is denoted by a yellow dotted square.

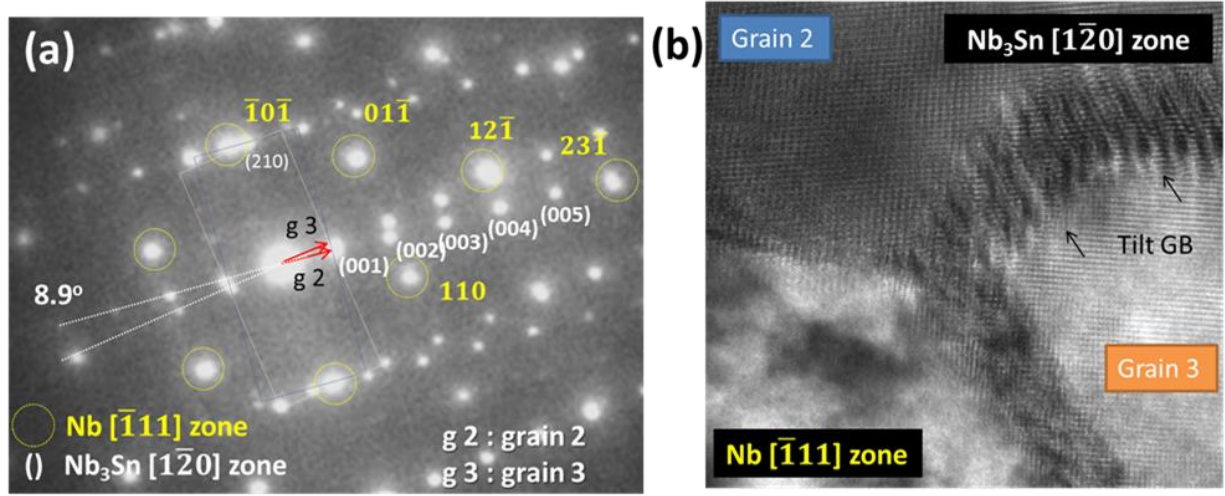

Figure 8. (a) Electron diffraction pattern of the $Nb_3Sn/Nb$ heterophase interface of grain numbers 2 and 3. The (002) plane of $Nb_3Sn$ grain number 2 is parallel to the $(23\bar{1})$ plane of Nb, denoted Orientation B. (b) HR-TEM image of the $Nb_3Sn/Nb$ interface of grain numbers 2 and 3, showing the formation of an $8.9°/[1\bar{2}0]$ tilt GB.

Grain number 5 was analyzed as a representative of the thin grains. HR-STEM imaging of the $Nb_3Sn/Nb$ interface with the thin-grain number 5 is displayed in Fig. 9(a) and it possess Orientation C, $Nb_3Sn$ $(1\bar{2}0)//Nb$ $(\bar{1}11)$ and $Nb_3Sn$ $(002)//Nb$ $(0\bar{1}1)$. For Orientation C, the $Nb_3Sn/Nb$ interface has a large lattice mismatch (12.3%) between the $Nb_3Sn$ (002) plane with d = 2.63 Å and the Nb $(0\bar{1}1)$ plane with d = 2.34 Å. Indeed, a high density of misfit dislocations, representing extra Nb $(0\bar{1}1)$ planes, is observed in the HR-STEM image, Fig. 9. The misfit dislocations appear at approximately every eight Nb $(0\bar{1}1)$ planes at the interface, which is in agreement with the lattice mismatch of 12.3%. A STEM-EDS map of the thin-grain in Fig. 9(b) reveals that an ~100 nm thick Sn-deficient layer exists at the $Nb_3Sn/Nb$ interface, probably due to the high compressive strain on the $Nb_3Sn$ $(1\bar{2}0)$ plane. The increase of the lattice parameter of $Nb_3Sn$ by adding Sn ($\frac{1}{a}\frac{da}{dc}$) is $\approx 2.6\times10^{-2}$ per at.% of Sn [36], which leads to a volume size factor ($\Omega_{sf}$) of Sn in Nb$_3$Sn of 8 % [37]. An HR-TEM study of another thin-grain (grain number 1) displays the same OR as the thin Nb$_3$Sn grain (grain number 5): see Fig. S.1 in Supplementary information. Orientations A, B, C, and D and their lattice mismatches are summarized in Table S.1. The lattice mismatches are estimated assuming stoichiometric 25 at.% Sn Nb$_3$Sn since the difference of the lattice parameter of Nb$_3$Sn between 17 to 25 at.% Sn is small, ~0.2% [36],

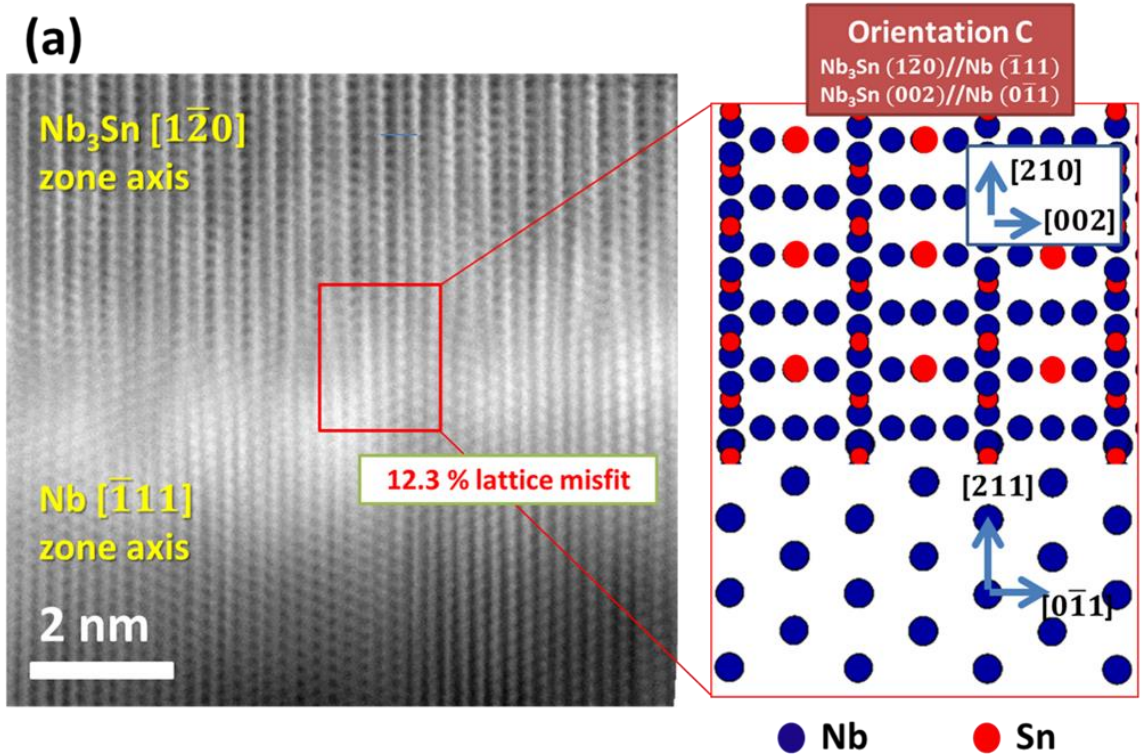

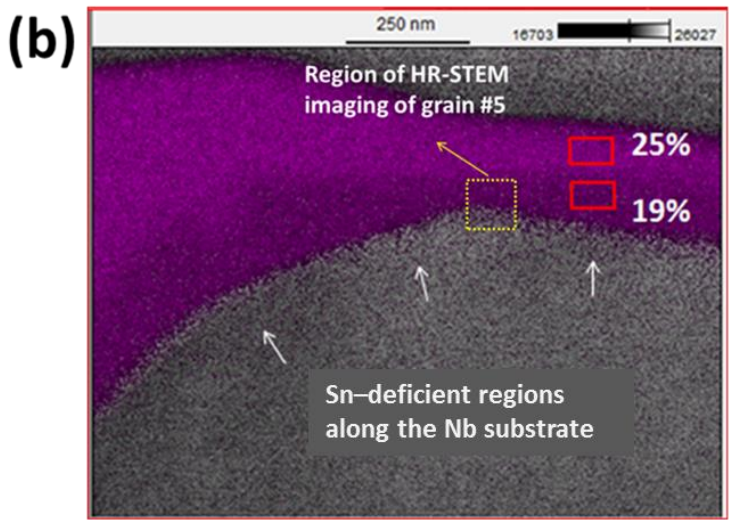

Figure 9. (a) Atomically-resolved HR-STEM image of the Nb₃Sn/Nb-heterophase interface of grain number 5 with Orientation C, and corresponding atomic configurations showing epitaxial growth of Nb₃Sn on Nb, right-hand side. The solid-red circles are Sn atoms and the solid-blue circles are Nb atoms. (b) HAADF-STEM EDS Sn L map of grain number 5. The region of the HR-STEM image at the Nb₃Sn (grain number 5)/Nb-heterophase interface is indicated by a yellow dotted square.

Transmission (t-) EBSD was performed to identify the misorientation axes and angles of Orientations A, B and C, Fig. 6. For Orientation A and B, both have an ~40° rotation angle about the <334> rotation axis between Nb₃Sn and Nb. They are not clearly distinguishable due to the small misorientation angle between Orientations A and B, 8.9° about the <120> axis, and deviations of the Euler angle measurements (~3°) determined from the EBSD patterns. Orientation C displays <123> or <124> rotation axes with ~50° of rotation, Table S.2. The misorientation axes and rotation angles of Orientations A, B and C were used to analyze the ORs of normal grain regions and patchy regions with thin-grains comparatively, using normal or transmission EBSD, section 4.2.

### 3.3 Vacancy formation and antisite substitutional behavior in Nb₃Sn

We used first-principles calculations to understand the formation of Nb and Sn antisite defects and vacancies. The vacancy formation energies, $E_V$, for both Nb and Nb₃Sn are described using a 2×2×2 supercell employing the following equation [38]:

$$E_V = E_V^{TOT} - E_{bulk}^{Total} + \mu \qquad (1)$$

Where $E_V^{TOT}$ is the total energy of the supercell with a vacancy, $E_{bulk}^{Total}$ is the total energy of the supercell without a vacancy, and $\mu$ is the chemical potential of the atom removed from the vacancy site. Our calculations yield the following results: (i) $E_V$ is 2.82 eV/atom for Nb in bulk Nb; (ii) 3.95 eV/atom for Sn in Nb$_3$Sn; and (iii) 2.04 eV/atom for Nb in Nb$_3$Sn. A Nb vacancy in Nb$_3$Sn has a smaller formation energy and thus is more readily formed than a Sn vacancy in Nb$_3$Sn. Both Nb and Sn vacancies play a major role for understanding bulk diffusion in Nb$_3$Sn.

The antisite substitutional structures were modeled by allowing either Nb or Sn to substitute at either the Sn or Nb sublattice sites in both the 2×2×2 and 3×3×3 Nb$_3$Sn superlattices and then fully relaxing the structures. The antisite energies in Nb$_3$Sn were calculated employing the following equations [38, 39]:

$$E_{Nb \to Sn} = (E^{TOT}_{Nb_3(Sn_{1-x}Nb_x)} + \mu_{Sn}) - (E^{TOT}_{Nb_3Sn} + \mu_{Nb}) \tag{2}$$

$$E_{Sn \to Nb} = (E^{TOT}_{(Nb_{1-y}Sn_y)_3Sn} + \mu_{Nb}) - (E^{TOT}_{Nb_3Sn} + \mu_{Sn}) \tag{3}$$

where $\mu_i$ is the chemical potential of Nb or Sn, $E^{TOT}_{Nb_3Sn}$ is the total energy of Nb$_3$Sn, $E^{TOT}_{Nb_3(Sn_{1-x}Nb_x)}$ is the total energy of Nb$_3$Sn with Nb at a Sn sublattice site, and $E^{TOT}_{(Nb_{1-y}Sn_y)_3Sn}$ is the total energy of Nb$_3$Sn with Sn in a Nb sublattice site. The first-principles results in Table 2 demonstrate that a Nb antisite atom forms more readily than a Sn antisite atom in both the 2×2×2 and 3×3×3 Nb$_3$Sn superlattices, because $E_{Nb \to Sn}$ is significantly smaller than $E_{Sn \to Nb}$. The Nb antisite atom generates both smaller average atomic forces and local atomic displacements. The experimental results in this study demonstrate the presence of Sn deficient regions close to the Nb substrate and the first-principle calculations indicate that Sn-deficient Nb$_3$Sn is an Nb-antisite compound, with Sn sites being replaced by Nb atoms, with Nb vacancies as secondary point defects with small concentrations of Sn vacancies due to their high vacancy formation energy.

Table 2. The antisite energies, average atomic force and displacement at the first nearest-neighbor distances associated with antisite substitutions, determined by first-principles calculations.

|  | Supercell | $E_{antisite}$ (eV atom$^{-1}$) | Average Atomic Force (eV Å$^{-1}$) | Average Atomic Displacement (Å) |
|---|---|---|---|---|
| $E_{Nb \rightarrow Sn}$ | 2×2×2 | 0.325 | 0.01674 | 0.0311 |
|  | 3×3×3 | 0.268 | 0.01251 | 0.0227 |
| $E_{Sn \rightarrow Nb}$ | 2×2×2 | 0.718 | 0.03118 | 0.0419 |
|  | 3×3×3 | 0.637 | 0.02886 | 0.0344 |

## 4. Discussion

### 4.1 Effect of Sn-flux (growth rate) on the microstructure of Nb$_3$Sn on Nb

This study describes the microstructure of three regions of the Nb$_3$Sn coatings on Nb, which are formed by different rates of a net Sn flux (or average growth rate). As Table 1 demonstrates, the microstructure of a Nb$_3$Sn coating is affected strongly by the net Sn-flux during its formation. This is important because a homogeneous high-quality Nb$_3$Sn coating on Nb with a reasonably smooth surface, no uncoated regions or thin grains, and smaller composition variations is critical for the performance of Nb$_3$Sn SRF cavities to avoid heating at microstructural imperfections. A uniform Nb$_3$Sn coating with an average grain size of 2 ± 0.6 μm and a thickness of ~2.5 μm was obtained with a medium Sn flux whose value is about 161 Sn atoms/nm$^2$ min, Fig. 4 and Table 1. Our results indicate that the Sn-flux has a strong influence on the kinetics of Nb$_3$Sn's nucleation and growth during the coating process. A large net Sn-flux (322 Sn atoms/nm$^2$ min), and a growth rate of 24 nm/min are associated with the formation of abnormally large grains (more than 5 μm) and a rough surface topology of a Nb$_3$Sn coating, Fig. 3. This is probably due to the abrupt formation of Nb$_3$Sn grains with the concomitant formation of liquid Sn-droplets on the surface. The melting temperature of Sn is low, 231.9 °C, so that Sn is in the liquid state at the process temperature, 1100 $^{o}$C. Therefore, if the flux of Sn atoms from the vapor phase is greater than the diffusion of Sn into bulk Nb$_3$Sn, Sn may accumulate on the surface and form liquid droplets on top of Nb$_3$Sn. This supposition is supported by the round morphology of the abnormally large grain regions representing Sn droplets formed in a range of diameters, from tens to hundreds of microns, when using a high Sn-flux, Fig. 3(a). In contrast, patchy regions with ~200 nm thick thin grains with a large lateral diameter more than ~4 micron, appear on the surface in the region with low net Sn-flux (47 Sn atoms/nm$^2$ min). This indicates that the nucleation of Nb$_3$Sn is non-uniform and some of the thin grains outgrow others in a lateral direction. Details of the origin of the thin Nb$_3$Sn grains are discussed in section 4.3.

### *4.2 Orientation relationships at Nb$_3$Sn/Nb*

In addition to the effect of the Sn-flux on the microstructure of Nb$_3$Sn, the primary finding is that Nb$_3$Sn/Nb heterophase interfaces play a critical role in the formation of imperfections in Nb$_3$Sn coatings, such as thin-grains and Sn-deficient regions. Specifically, a strong correlation exists between the formation of thin-grains and Orientation C, Fig. 9. Four types of ORs for the Nb$_3$Sn/Nb heterophase interfaces (Orientations A, B, C and D) were found and three of these ORs (Orientation A, B, C) were frequently observed on the zone axis of Nb [$\bar{1}$11] with Nb$_3$Sn [1$\bar{2}$0], Fig. 6, which suggests that these interfaces have lower interfacial free energies than alternative interfaces.

It is also noteworthy that the orientation relationships, in particular, Orientations A, B and C, were observed more frequently in patchy regions for the case of a low Sn-flux compared to medium and high Sn-fluxes. EBSD and transmission EBSD analyses were employed to analyze ORs of Nb$_3$Sn/Nb heterophase interfaces in normal grain-regions and patchy regions with thin grains, Fig. 10 and Table 3. A total 66 interfaces of normal grain regions and 13 interfaces of patchy regions with thin grains were analyzed and the latter display a higher frequency of Orientations A, B and C, ~ 69%, compared to the interfaces in normal grain regions, ~17%, Table 3. This is probably because the small Sn-flux results in slow growth of $Nb_3Sn$ grains, providing sufficient time for the Sn atoms to diffuse into the Nb substrate and form a stable $Nb_3Sn$/Nb interface.

Table 3. Statistics of orientation relationships at $Nb_3Sn$/Nb interfaces in normal $Nb_3Sn$ regions and patchy regions with thin $Nb_3Sn$ grains. Sixty-six grains were analyzed in the normal regions and thirteen grains were analyzed in the patchy regions with thin thin-grains using EBSD, transmission EBSD, and TEM analyses.

| Orientation type | Frequencies | |
|---|---|---|
| | Normal $Nb_3Sn$ regions (N = 66) | Patchy regions with thin $Nb_3Sn$ grains (N=13) |
| Orientation A | 5% [a] | 15 % |
| Orientation B | | 15 % |
| Orientation C | 12 % | 39 % |
| Others | 83 % | 31 % |

[a] Orientation A and B are not distinguishable in transmission (t-) EBSD due to the deviations of the Euler angles in the EBSD data (see Table S.2 in Supplementary information).

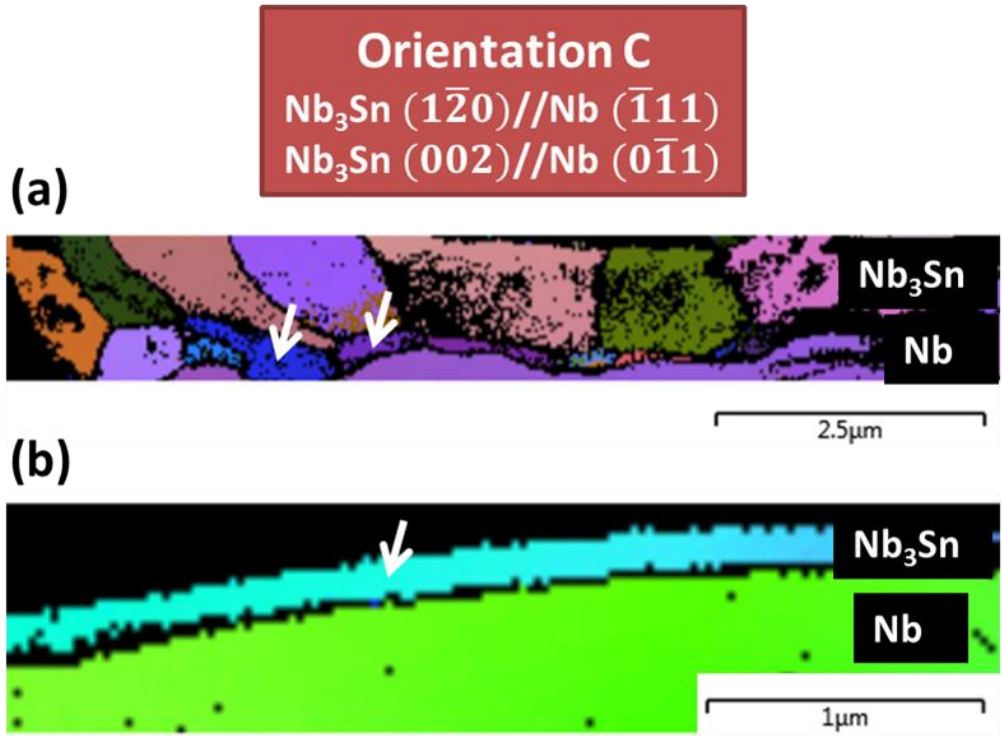

Figure 10. Transmission EBSD map of: (a) normal grain regions and (b) patchy regions with thin-grains. Two grains among fifteen grains in the TEM sample with normal regions (a) have Orientation C, denoted by white arrows. Also, the thin-grain (b) has Orientation C. Transmission electron-microscope sample (a) was prepared at Fermilab and sample (b) was prepared at Cornell University.

### *4.3 Origin of the formation of thin grains*

Grains with Orientation C [$Nb_3Sn$ $(1\bar{2}0)$//Nb $(\bar{1}11)$ and $Nb_3Sn$ $(002)$//Nb $(0\bar{1}1)$] are consistently significantly thinner than the other grains, indicating that the interfacial migration rate of these grains is slower than that of the other grains. The slow interfacial migration rate can be rationalized by the high dislocation density of misfit dislocations due to a large lattice mismatch, 12.3%. The HR-STEM image, Fig. 9, demonstrates that the misfit dislocations are separated by ~2 nm. The migration of these interfaces requires climbing and/or gliding of the misfit dislocations and the energy required for these processes could retard the velocity of the reaction front [40, 41].

The influence of ORs on the interfacial migration rate, during a solid-state reaction, has been reported for other systems, especially growth of a spinel on MgO or sapphire ($Al_2O_3$) [40-48]. For $MgAl_2O_4$ spinel-growth between MgO and $Al_2O_3$, the $MgAl_2O_4$/MgO interface is pinned by a high misfit dislocation-density; one dislocation every ~23 atomic plane of (010) $MgAl_2O_4$, caused by the large lattice mismatch ($2\frac{a_{MgAl2O4}- 2a_{MgO}}{a_{MgAl2O4}+ 2a_{MgO}} \approx$ -4.3%) [47]. Similarly, for $ZnAl_2O_4$ growth between ZnO and $Al_2O_3$, the reaction at the $ZnAl_2O_4$/ZnO interface, which has a large lattice mismatch (-13.7%), is limited due to dislocation gliding and/or climbing, while the $ZnAl_2O_4$/$Al_2O_3$ interface with a smaller lattice mismatch (2.1%) is relatively mobile [45]. Interfacial migration of an interface with a large lattice mismatch requires climb of the misfit dislocations at the interface, which is the rate-limiting step [45, 47]. In the case of $Nb_3Sn$, the nucleated grain with Orientation C may preferentially spread in the lateral direction rather than increasing its thickness. This would give rise to the formation of abnormally large thin-grains during the early stage of $Nb_3Sn$ growth. The formation of large thin-grains significantly reduces the Sn supply to the $Nb_3Sn$/Nb interface due to the decreased density of grain boundaries (which act as short circuit diffusion pathways). As a result, the interfacial reaction rate at $Nb_3Sn$/Nb becomes even slower.

Another factor that influences the formation of thin $Nb_3Sn$ grains is the growth rate [49]. Thin grains with a large lateral diameter appeared when the average growth rate was slow (~3.5 nm/min) as a consequence of a small net Sn flux (~47 Sn atoms/$nm^2$ min), and they are not observed in medium and high net Sn-flux regions. The experimental evidence in this article is not sufficient to demonstrate that the average tin flux during the coating process is represented by the net tin-flux measured after coating, but it is logical that a high average tin-flux could help prevent the growth of large thin grains. When the Sn flux is high, the density of nucleation sites increases and, therefore, the lateral growth of the thin grains is limited due to the competition with neighboring $Nb_3Sn$ grains.

*4.4 Formation of Sn-deficient regions: nucleation and their evolution*

In the Nb-Sn binary phase diagram, Fig.1, the composition of $Nb_3Sn$ in the two-phase equilibrium phase-field ($\alpha$−Nb plus $Nb_3Sn$) is ~17 at.% Sn at 1100 $^oC$, and it is therefore reasonable that the nucleated $Nb_3Sn$ grains at the $Nb_3Sn$/Nb heterophase interface have this Sn concentration, Fig. 2. As discussed, the Sn-deficient regions are probably initially formed at the $Nb_3Sn$/Nb-heterophase-interface and the Sn deficiency in the middle of a grain is a consequence of slow Sn diffusion, ~100 nm/h [50], in $Nb_3Sn$. First-principle calculations revealed that the Nb anti-site defect has a small formation energy, ~0.3 eV/atom (Table 2) and the formation of Sn-deficient regions may be assisted by the small formation energy of Nb anti-site defects [26, 34, 50]. The concentration of Sn-deficient regions is 17 to 19 at.% Sn, and this implies that 0.24 to 0.32 fraction of Sn sites are occupied by Nb atoms in these regions.

The interfacial energy of the Nb/$Nb_3Sn$ heterophase-interface is essential for understanding the mechanisms and stability of this heterophase-interface with vicinal orientations. Specifically, the interfacial energy plays a critical role in the nucleation of $Nb_3Sn$ because the strain energy and the formation energy of Nb antisite defects are negligible compared to the interfacial energy in the nucleation stage [51-53]. We calculate the energies of the four simple interfaces with Orientations A and B, which our experiments detected: (i) $Nb_3Sn$ ($1\bar{2}0$)//Nb($\bar{1}11$); (ii) $Nb_3Sn$(002)//Nb($1\bar{1}2$); (iii) $Nb_3Sn$(210)//Nb(110) with Orientation A; and (iv) $Nb_3Sn$(002)//Nb($23\bar{1}$) with Orientation B. Orientation C is excluded because it exhibits abnormal $Nb_3Sn$ growth and difficult to simulate using first-principles calculations due to the large lattice mismatch. The interfacial configurations were constructed with the initial positions based on ideal Nb and $Nb_3Sn$ phases; the latter was stretched to match the Nb phase. We subsequently fully relaxed all the atomic positions, with the unit vector normal to the (hkl) plane of the interface. The interfacial internal energies at 0 K are calculated by subtracting the total energy of the phases on either side of the interface from the total energy of a two-phase system containing an interface:

$$\sigma_{Nb/Nb_3Sn} = \frac{1}{2A}\left[E^{tot}_{Nb/Nb_3Sn} - (E^{tot}_{Nb} + E^{tot}_{Nb_3Sn})\right] \quad (4)$$

where $A$ is the interfacial area, $E_{Nb}^{tot}$ is the total internal energy of the Nb phase, $E_{Nb_3Sn}^{tot}$ is the total internal energy of the stretched Nb$_3$Sn phase, $E_{Nb/Nb_3Sn}^{tot}$ is the total internal energy of the relaxed Nb/Nb$_3$Sn system containing an interface; this was done for four different orientations. The calculated values of the interfacial internal energies at 0 K are summarized in Table 4. Nb$_3$Sn(002)//Nb($1\bar{1}2$) with Orientation A has the lowest interfacial internal energy and Nb$_3$Sn(210)//Nb(110) with Orientation A has the largest interfacial internal energy. We note that the interfaces with Sn-deficient Nb$_3$Sn for all four orientations have 11 to 12% smaller interfacial internal energies than those with the exact Nb$_3$Sn stoichiometry, Table 4. In the early nucleation stage, the change of the bulk internal energy is negligible and Nb$_3$Sn nucleates readily in regions with a small interfacial internal energy. This reflects the fact that the nucleated Nb$_3$Sn, Fig. 2, is Sn-deficient and there is a large amount of nucleated Sn-deficient Nb$_3$Sn existing at Nb$_3$Sn/Nb heterophase-interfaces in a Nb$_3$Sn layer in our experiments, Figs. 4, 7 and 9.

Once the Sn-deficient Nb$_3$Sn phase is nucleated and grows at Nb$_3$Sn/Nb-heterophase-interfaces, Sn-deficient Nb$_3$Sn transforms to stoichiometric Nb$_3$Sn by replacing Nb antisite atoms with Sn. Diffusion of Sn within a Nb$_3$Sn grain is slow due to the highly correlated diffusion of Sn in this ordered structure [26, 34, 50]. We note that the diffusion coefficient of Sn in Nb$_3$Sn is approximately three orders of magnitude smaller than even that of Nb at 1100 $^o$C and the root-mean-square diffusion distance of Sn in Nb$_3$Sn is ~100 nm in one hour at 1100 $^o$C [50, 54]. As a consequence, a coating time of 3.5 h is insufficient to allow Sn-deficient regions to achieve the equilibrium concentration of 25 at.% Sn, if the size of the Sn-deficient regions is more than ~350 nm.

Based on our observations and prior studies [18, 19, 35], we propose a scheme for the growth of normal Nb$_3$Sn grains on Nb that is summarized in Fig. 11. At the nucleation step, Fig.11, Sn vapor condenses on the Nb surface forming Sn nucleation sites for Nb$_3$Sn embryos, which become stable Nb$_3$Sn nuclei that eventually become Nb$_3$Sn grains. According to the phase diagram, first-principles calculations, and STEM-EDS results, the Nb$_3$Sn nuclei are likely to be Sn deficient, see Figs. 1, 2 and Table 4. When the temperature is increased to 1100 °C, significant grain growth of $Nb_3Sn$ occurs [18] and the lateral growth of $Nb_3Sn$ results in impingement and coalescence of $Nb_3Sn$ grains, which result in grain boundaries [55]. Once $Nb_3Sn$ covers completely the Nb substrate, the growth of the $Nb_3Sn$ coating is controlled by an interfacial reaction at the $Nb_3Sn$/Nb heterophase interface and is limited by the diffusion of Sn along the grain-boundaries, that is, short-circuit diffusion. In this case, there could be Sn-deficient regions in the middle of grains resulting from slow Sn-diffusion in $Nb_3Sn$, which is highly correlated.

Table 4. The interfacial internal energy of four different orientations at 0 K calculated using first-principles calculations, Unit: $mJ/m^2$

| Orientation relationship | Interface type | Perfect $Nb_3Sn$ | Sn-deficient[a] $Nb_3Sn$ |
|---|---|---|---|
| A | $Nb_3Sn$ $(1\bar{2}0)$//Nb $(\bar{1}11)$ | 262 | 229 |
|   | $Nb_3Sn$ $(002)$//Nb $(1\bar{1}2)$ | 208 | 186 |
|   | $Nb_3Sn$ $(210)$//Nb $(110)$ | 274 | 233 |
| B | $Nb_3Sn$ $(002)$//Nb $(23\bar{1})$ | 223 | 195 |

[a.] The interfacial structure is Nb/$Nb_3Sn$. The Nb slab has 66 atoms and the $Nb_3Sn$ slab has 64 atoms. Three Nb antisites are employed in the Sn-deficient $Nb_3Sn$ slab resulting in 20.3 at.% Sn.

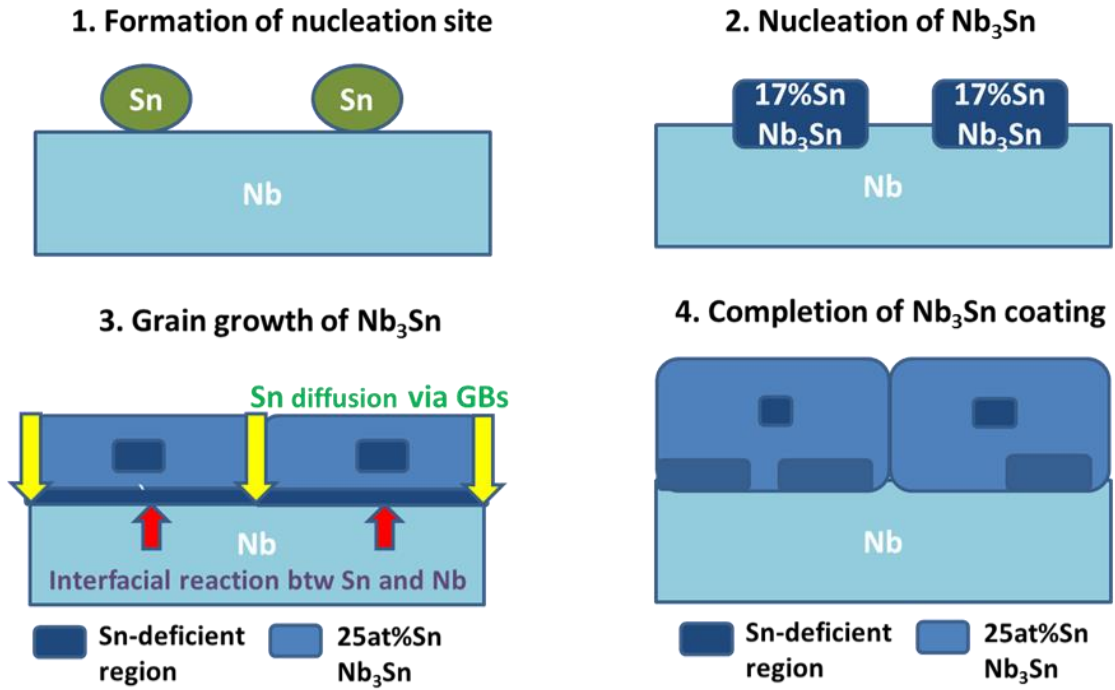

Figure 11. Schematic diagram for proposed coating process of Nb$_3$Sn on Nb employing vapor diffusion of Sn. The nucleated embryos of Nb$_3$Sn in step 2 (Nucleation of Nb$_3$Sn) grow laterally to form a continuous layer of Nb$_3$Sn, which has regions that are deficient in Sn (step 3 – grain growth of Nb$_3$Sn). In step 4 (formation of a complete layer of Nb$_3$Sn), note the presence of a grain boundary, which is an important short-circuit for diffusion of Sn toward the Nb$_3$Sn/Nb heterophase interface.

### 4.5 Correlation between microstructure and superconductivity of Nb$_3$Sn superconducting radiofrequency (SRF) cavities

Ref. [14] describes the study in which a direct correlation was made between the degradation of the Q$_0$ of an SRF cavity (Q vs E curve of ERL1-5 in Fig. 1) and thin-grains in the regions of degraded performance. We reexamined one of the coupons from a region with degraded performance and analyzed one of the thin-grains for ORs. It also exhibited Orientation C as demonstrated using electron diffraction, Fig. 12 (c,d), which provides additional strong evidence that the degradation of superconductivity in a Nb$_3$Sn cavity is due to thin-grains, which is caused by specific ORs of Nb$_3$Sn/Nb associated with small net Sn-fluxes. Therefore, a critical net Sn flux is required to avoid the formation of thin-grains with Orientation C. Also, pre-anodization of Nb substrates, which introduces 70~100 nm niobium oxide layers on Nb, is employed to induce homogeneous nucleation of $Nb_3Sn$ embryos at a high number density, which significantly reduces the large lateral growth of the thin-grains with Orientation C, Fig. 12(a,b) [18, 35]. Indeed, the grain with Orientation C found in the $Nb_3Sn$ coating on pre-anodized Nb is similar in grain size (~ 2 μm) to other neighboring grains, Fig. 12(a,b). Furthermore, the cavity with the grain with Orientation C exhibits no degradation of superconductivity until 17 MV/m, similar to ERL 1-4 in Fig. 1(b). This demonstrates that even though grains with Orientation C may be present in a $Nb_3Sn$ coating, the detrimental effects of thin grains on SRF cavity performance can be mitigated by controlling the Sn-flux and inducing homogeneous nucleation of $Nb_3Sn$ embryos using pre-anodization of Nb substrates. It is possible that Nb oxide-layers on the surface may interfere with the epitaxial growth of $Nb_3Sn$ on Nb, causing more randomly oriented and more chemically uniform $Nb_3Sn$ coatings.

The effect of Sn-deficient $Nb_3Sn$ on the superconductivity of $Nb_3Sn$ SRF cavity is anticipated to be unfavorable because the $T_c$ of Sn-deficient $Nb_3Sn$ (17 at.% Sn) decreases from 18.3 K of perfect $Nb_3Sn$ to 6 K, which is less than the $T_c$ of Nb (9.2 K). The distribution of Sn-deficient regions in the $Nb_3Sn$ coatings and detailed correlations between Sn-deficient regions and SRF cavity performance require additional in-depth studies.

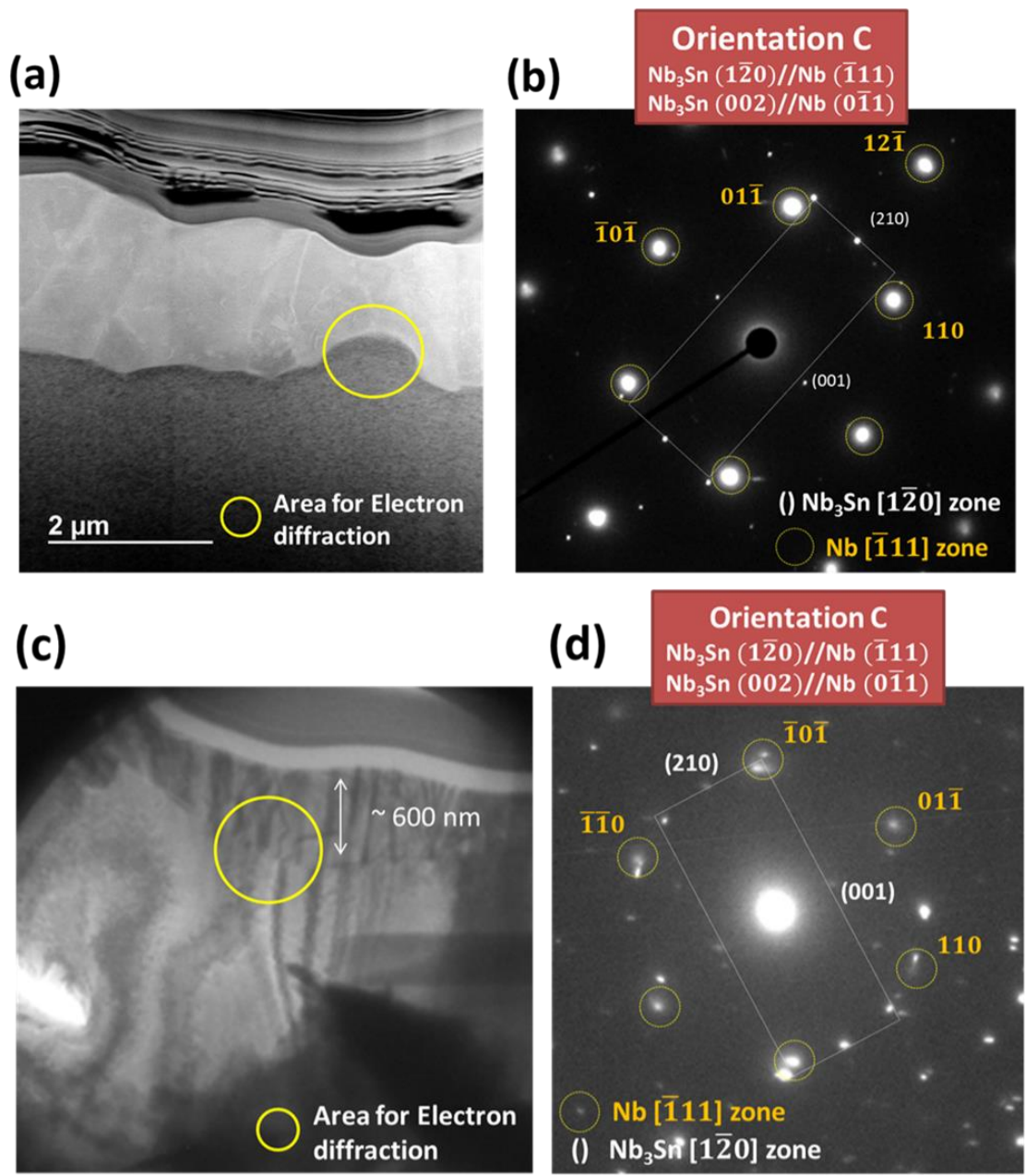

Figure 12. (a) HAADF-STEM image and (b) electron diffraction pattern of the relatively thinner $Nb_3Sn$ grain (~700 nm) compared to other neighboring grains (~2 μm) in a good performing SRF cavity similar to ERL 1-4. (c) BF-TEM image of a thin-grain from the performance-degraded region of ERL 1-5 $Nb_3Sn$ cavity, and (d) corresponding electron diffraction pattern. Both $Nb_3Sn$ films were prepared at Cornell University.

**Conclusions**

Nb$_3$Sn coatings on Nb prepared by a vapor-diffusion process for superconducting radio frequency (SRF) cavity applications were analyzed systematically using transmission electron microscopy (TEM), electron backscatter diffraction (EBSD) and first-principles calculations.

- Four types of orientation relationships (ORs) between Nb$_3$Sn and Nb (Orientations A, B, C and D) were detected and analyzed by electron diffraction and high-resolution scanning transmission electron microscopy (HR-STEM). Notably, there is a close relationship between the formation of thin-grains in low Sn-flux regions and the ORs, Nb$_3$Sn ($1\bar{2}0$)//Nb ($\bar{1}11$) and Nb$_3$Sn (002)//Nb ($0\bar{1}1$), termed Orientation C. Orientation C displays a large lattice mismatch (12.3%) between Nb$_3$Sn (002) and Nb ($0\bar{1}1$) and a high density of misfit dislocations as observed employing HR-STEM images.

- The formation of abnormally thin-grains is attributed to the slow migration of the Nb$_3$Sn/Nb heterophase interface with Orientation C, caused by its high density of misfit dislocations.

- The formation of Sn-deficient regions at the heterophase interfaces were also quantified employing first-principle calculations. We found that heterophase interfaces with Sn-deficient Nb$_3$Sn have smaller interfacial internal energies than those with perfect Nb$_3$Sn.

- In the early nucleation stage, the new phase with a small interfacial free energy nucleates readily when the change of the bulk internal energy is small, which results in large areas of Sn-deficient Nb$_3$Sn existing in the Nb$_3$Sn layer as experimentally demonstrated.

- The Nb$_3$Sn/Nb heterophase interfaces and the crystallographic orientation relationships of Nb$_3$Sn with Nb play important roles in the formation of abnormal thin-grains and Sn-deficient regions, imperfections that are detrimental to the superconducting properties of Nb$_3$Sn SRF cavities.

- The current study may yield new possibilities for controlling the imperfections in Nb$_3$Sn coatings and improving their quality to increase the accelerating electric field of Nb$_3$Sn cavities.

**Acknowledgements**

We are grateful to Drs. Amir R. Farkoosh, Xuefeng Zhou, and Sung-Il Baik, and Mr. Qingqiang Ren for valuable discussions. This research is supported by the United States Department of Energy, Offices of High Energy. Fermilab is operated by the Fermi Research Alliance LLC under Contract No. DE-AC02-07CH11359 with the United States Department of Energy. This work made use of the EPIC, Keck-II, and/or SPID facilities of Northwestern University's NU*ANCE* Center, which received support from the Soft and Hybrid Nanotechnology Experimental (SHyNE) Resource (NSF ECCS-1542205); the MRSEC program (NSF DMR-1121262) at the Materials Research Center; the International Institute for Nanotechnology (IIN); the Keck Foundation; and the State of Illinois, through the IIN. Cornell's $Nb_3Sn$ coating program is supported by United States Department of Energy grant DE-SC0008431. NUCAPT received support from the MRSEC program (NSF DMR-1720139) at the Materials Research Center, the SHyNE Resource (NSF ECCS-1542205), and the Initiative for Sustainability and Energy (ISEN) at Northwestern University.